\documentclass[12pt]{article}
\pdfoutput=1
\usepackage{subfigure}
\usepackage{amssymb,amsmath}
\usepackage{graphicx}
\usepackage{color}
\usepackage[colorlinks=true
,urlcolor=blue
,citecolor=blue
,linkcolor=blue
,pagecolor=blue
,linktocpage=true
,pdfproducer=medialab
]{hyperref}
\usepackage[a4paper,width=15.2cm]{geometry}
\makeatletter \renewcommand{\@dotsep}{10000} \makeatother

\newcommand{\beq}{\begin{equation}}
\newcommand{\eeq}{\end{equation}}
\newcommand{\bea}{\begin{eqnarray}}
\newcommand{\eea}{\end{eqnarray}}

\begin{document}

\begin{center}

 {\Large\bf   Higgs Boson  Mass  from $t$-$b$-$\tau$ Yukawa Unification
 } \vspace{1cm}

{\large   Ilia Gogoladze\footnote{E-mail: ilia@bartol.udel.edu\\
\hspace*{0.5cm} On  leave of absence from: Andronikashvili Institute
of Physics, 0177 Tbilisi, Georgia.}, Qaisar Shafi\footnote{ E-mail:
shafi@bartol.udel.edu} and Cem Salih $\ddot{\rm U}$n \footnote{
E-mail: cemsalihun@bartol.udel.edu}} \vspace{.5cm}

{\baselineskip 20pt \it
Bartol Research Institute, Department of Physics and Astronomy, \\
University of Delaware, Newark, DE 19716, USA  } \vspace{.5cm}

\vspace{1.5cm}
 {\bf Abstract}
\end{center}

We employ the Yukawa coupling unification condition, $y_t= y_b= y_{\tau}$  at $M_{\rm GUT}$, inspired by supersymmetric SO(10) models, to estimate the lightest Higgs boson mass as well as masses of the associated squarks and gluino. We employ non-universal soft masses, dictated by SO(10) symmetry, for the gauginos. Furthermore, the soft masses for the two scalar Higgs doublets are set equal at $M_{\rm GUT}$, and in some examples these are equal to the soft masses for scalars in the matter multiplets. {For $\mu > 0$, $ M_{2} > 0 $, where $ M_{2} $ denotes the $ SU(2) $ gaugino mass, essentially perfect $ t$-$b$-$\tau $ Yukawa unification is possible, and it predicts a Higgs mass of 122 - 124 GeV with a theoretical uncertainty of about $ \pm 3 $ GeV.} The corresponding gluino and the first two family squarks have masses $\gtrsim 3$ TeV.
We present some LHC testable benchmark points which also show the presence of neutralino-stau coannihilation in this scenario. The well-known MSSM parameter   $ \tan\beta\approx 47 $.

\newpage

\renewcommand{\thefootnote}{\arabic{footnote}}
\setcounter{footnote}{0}



\section{\label{ch:introduction}Introduction}

Supersymmetric (SUSY) $SO(10)$ grand unified theory (GUT), in
contrast to its non-SUSY version, yields third family
($t$-$b$-$\tau$) Yukawa unification via the unique renormalizable
Yukawa coupling $16 \cdot 16 \cdot 10$, if the Higgs 10-plet is assumed
to contain the two Higgs doublets $H_u$ and $H_d$ of the minimal supersymmetric standard model (MSSM) \cite{big-422}. The matter 16-plet contains the 15
chiral superfields of MSSM as well as the
right handed neutrino superfield. The implications of this Yukawa unification condition at $ M_{\rm GUT} \sim 2\times 10^{16}$ GeV have
been extensively explored over the years \cite{big-422,bigger-422}.
In $SO(10)$ Yukawa unification with $ \mu > 0 $ and universal gaugino masses, the gluino is the lightest colored sparticle \cite{Baer:2008jn,
Gogoladze:2009ug}, which is now being tested \cite{Baer:2009ff} at the Large Hadron Collider (LHC). The
squarks and sleptons, especially those from the first two families,  turn out to have masses in the multi-TeV
range. {Moreover, it is argued in \cite{Baer:2008jn,
Gogoladze:2009ug}, based on the results of publicly available codes like Isajet \cite{ISAJET} and Softsusy \cite{Allanach:2001kg}}, that the lightest neutralino is not a viable cold
dark matter candidate in $SO(10)$
Yukawa unification with $ \mu > 0 $ and universal gaugino masses at $ M_{\rm GUT} $.
{On the other hand, in Ref.[3] it is shown that neutralino DM is allowed through h-resonance for Yukawa unification at the 10$\% $level.}

Spurred by these developments we have investigated $t$-$b$-$\tau$
Yukawa unification \cite{Gogoladze:2009ug,
Gogoladze:2009bn,Gogoladze:2010fu} in the framework of
SUSY  $SU(4)_c \times SU(2)_L \times SU(2)_R$ \cite{pati}
(4-2-2, for short). The 4-2-2 structure allows us to consider
non-universal gaugino masses while preserving Yukawa unification. An
important conclusion reached in \cite{Gogoladze:2009ug,
Gogoladze:2009bn} is that with same sign non-universal gaugino soft
terms, Yukawa unification in 4-2-2 is compatible with neutralino
dark matter and gluino co-annihilation \cite{Gogoladze:2009ug,Baer:2009ff,
Gogoladze:2009bn, Profumo:2004wk} being a unique dark matter scenario for $ \mu > 0 $.

By considering opposite sign gauginos with
{$\mu<0,M_2<0,M_3>0$} where $\mu$ is the coefficient of the bilinear Higgs mixing
term, $M_2$ and $M_3$ are the soft supersymmetry breaking (SSB)
gaugino mass terms corresponding  to $SU(2)_L$ and
$SU(3)_c$ respectively, it is shown in \cite{Gogoladze:2010fu} that Yukawa
coupling unification consistent with the experimental constraints
can be implemented in 4-2-2. With $\mu<0$ and opposite sign gauginos,
Yukawa coupling unification is achieved for $m_{16} \gtrsim 300\, {\rm
GeV}$, as opposed to $m_{16} \gtrsim 8\, {\rm TeV}$ for the case of
same sign gauginos. The finite corrections to the b-quark
mass play an important role here \cite{Gogoladze:2010fu}. By considering gauginos with $M_2 <0$, $M_3>0$ and $\mu<0$,
we can obtain the correct sign for the desired contribution to
$(g-2)_\mu$ \cite{Bennett:2006fi}. This enables us to simultaneously
satisfy the requirements of  $t$-$b$-$\tau$ Yukawa unification in 4-2-2,
neutralino dark matter and $(g-2)_\mu$, as well as a variety of
other bounds.

Encouraged by the abundance of solutions and co-annihilation channels
available in the case of Yukawa unified 4-2-2 with $M_2 <0$ and
$\mu<0$, it seems natural to explore Yukawa unification in SO(10) GUT (with $M_2 <0$ and $\mu<0$). It has been pointed out \cite{Martin:2009ad} that non-universal MSSM gaugino masses at $ M_{\rm GUT} $ can arise from non-singlet F-terms, compatible with the underlying GUT symmetry such as SU(5) and SO(10). The SSB
gaugino masses in supergravity  \cite{Chamseddine:1982jx} can arise, say, from the following
dimension five operator:
\begin{align}
 -\frac{F^{ab}}{2 M_{\rm
P}} \lambda^a \lambda^b + {\rm c.c.}
\end{align}
 Here $\lambda^a$ is the two-component gaugino field, $ F^{ab} $ denotes the F-component of the field which breaks SUSY, the indices $a,b$ run over
the adjoint representation of the gauge group, and  $M_{P}=2.4 \times 10^{18}$ GeV is the reduced Planck mass.
  The resulting gaugino
mass matrix is $\langle F^{ab} \rangle/M_{\rm P}$ where the
supersymmetry breaking  parameter $\langle F^{ab} \rangle$
transforms as a singlet under the MSSM gauge group $SU(3)_{c}
\times SU(2)_L \times U(1)_Y$. The $F^{ab}$ fields belong to an
irreducible representation in the symmetric part of the direct product of the
adjoint representation of the unified group. This is a supersymmetric generalization of operators considered a long time ago \cite{Hill:1983xh}.

In SO(10), for example,
\begin{align}
({ 45} \times { 45} )_S = { 1} + { 54} + { 210} +
{ 770}
\end{align}
If  $F$  transforms as a 54 or 210 dimensional
representation of SO(10) \cite{Martin:2009ad}, one obtains the following relation
among the MSSM gaugino masses at $ M_{\rm GUT} $ :
\begin{align}
M_3: M_2:M_1= 2:-3:-1 ,
\label{gaugino10}
\end{align}
where $M_1, M_2, M_3$ denote the gaugino masses of $U(1)$, $SU(2)_L$ and $SU(3)_c$
respectively. The low energy implications of this relation have recently been investigated in \cite{Okada:2011wd} without imposing Yukawa unification. We consider the case with $ \mu > 0 $ and non-universal gaugino masses defined in Eq.(\ref{gaugino10}) in this paper. In order to obtain the correct sign for the desired contribution to $ (g-2)_{\mu} $, we set signs of the gaugino masses as $ M_{1} > 0 $, $ M_{2} > 0 $ and $ M_{3} < 0 $. Somewhat to our surprise, we find that this class of $ t$-$b$-$\tau $ Yukawa unification models make a rather sharp prediction for the mass of the lightest SM-like Higgs boson. In addition, lower mass bounds on the masses of the squarks and gluino are obtained.

{Notice that in  general, the soft terms such as the trilinear couplings and scalar (mass)$^2$ terms are not necessarily universal at $M_{GUT}$}. However, we can assume, consistent with SO(10) gauge symmetry, that the coefficients associated with terms that violate the SO(10)-invariant form are suitably small, {except for} the gaugino term in Eq.(1). {We also assume that D-term contributions to the SSB term are much smaller compared with contributions from fields with non-zero auxiliary F-terms.}

The outline for the rest of the paper is as follows.
In Section \ref{constraintsSection} we summarize the scanning procedure and the
experimental constraints that we have employed. In Section \ref{muneg} we discuss the important role of SUSY threshold corrections in $ t$-$b$-$\tau $ Yukawa unification. In Section \ref{huhd} we discuss how radiative electroweak symmetry breaking (REWSB) is compatible in our model despite setting $ m_{H_{u}}=m_{H_{d}} $ at $ M_{\rm GUT} $. In Section \ref{results} we present our results, focusing in particular on the mass of the lightest (SM-like) CP even Higgs boson, as well as the masses of squarks and gluino. The table in this section lists some benchmark points which will be tested at the LHC. Our conclusions are summarized in Section \ref{conclusions}.

\section{Phenomenological Constraints and Scanning Procedure\label{constraintsSection}}

We employ the ISAJET~7.80 package~\cite{ISAJET}  to perform random
scans over the fundamental parameter space. In this package, the weak scale values of gauge and third generation Yukawa
couplings are evolved to $M_{\rm GUT}$ via the MSSM renormalization
group equations (RGEs) in the $\overline{DR}$ regularization scheme.
We do not strictly enforce the unification condition $g_3=g_1=g_2$ at $M_{\rm
GUT}$, since a few percent deviation from unification can be
assigned to unknown GUT-scale threshold
corrections~\cite{Hisano:1992jj}.
The deviation between $g_1=g_2$ and $g_3$ at $M_{\rm GUT}$ is no
worse than $3-4\%$.
For simplicity  we do not include the Dirac neutrino Yukawa coupling
in the RGEs, whose contribution is expected to be small.

The various boundary conditions are imposed at
$M_{\rm GUT}$ and all the SSB
parameters, along with the gauge and Yukawa couplings, are evolved
back to the weak scale $M_{\rm Z}$.
In the evaluation of Yukawa couplings the SUSY threshold
corrections~\cite{Pierce:1996zz} are taken into account at the
common scale $M_{\rm SUSY}= \sqrt{m_{{\tilde t}_L}m_{{\tilde t}_R}}$
where $m_{{\tilde t}_L}$ and $m_{{\tilde t}_R}$
are the third generation left and right handed stop quarks.
 The entire
parameter set is iteratively run between $M_{\rm Z}$ and $M_{\rm
GUT}$ using the full 2-loop RGEs until a stable solution is
obtained. To better account for leading-log corrections, one-loop
step-beta functions are adopted for gauge and Yukawa couplings, and
the SSB parameters $m_i$ are extracted from RGEs at multiple scales
$m_i=m_i(m_i)$. The RGE-improved 1-loop effective potential is
minimized at $M_{\rm SUSY}$, which effectively
accounts for the leading 2-loop corrections. Full 1-loop radiative
corrections are incorporated for all sparticle masses.

The approximate error of  $\pm 3$ GeV in the ISAJET estimate of the
Higgs mass
largely arise from theoretical  uncertainties  in the calculation of the minimum of the scalar potential,  and
to a lesser extend from experimental uncertainties in the values
for $m_t$ and $\alpha_s$.

An important constraint comes from limits on the cosmological abundance of stable charged
particles  \cite{Nakamura:2010zzi}. This excludes regions in the parameter space
where  charged SUSY particles  become
the lightest supersymmetric particle (LSP). We accept only those
solutions for which one of the neutralinos is the LSP and saturates
the WMAP  bound on relic dark matter abundance.

The MSSM Higgs doublets reside in the 10 dimensional representation of $ SO(10) $ and fermions of the third  family belong to the 16 dimensional representation
of SO(10), which implies Yukawa coupling unification at $ M_{\rm GUT} $.

We have performed random scans for the following parameter range:
\begin{align}
0\leq  m_{16}  \leq 5\, \rm{TeV} \nonumber \\
0\leq   m_{10} \leq 5\, \rm{TeV} \nonumber \\
0 \leq M_{1/2}  \leq 2 \, \rm{TeV} \nonumber \\
35\leq \tan\beta \leq 55 \nonumber \\
-3\leq A_{0}/m_{16} \leq 3 \\
 \label{parameterRange}
\end{align}
 Here $ m_{16} $ is the universal SSB mass for MSSM sfermions, $ m_{10} $ is the universal SSB mass term for up and down MSSM Higgs masses, $ M_{1/2} $ is the gaugino mass parameter, $ \tan\beta $ is the ratio of the vacuum expectation values (VEVs) of the two MSSM Higgs doublets, $ A_{0} $ is the universal SSB trilinear scalar interaction (with corresponding Yukawa coupling factored out). We use    $m_t = 173.1\, {\rm GeV}$  \cite{:1900yx}. Note that our results are not
too sensitive to one or two sigma variation in the value of $m_t$  \cite{Gogoladze:2011db}.
We use $m_b(m_Z)=2.83$ GeV which is hard-coded into ISAJET.

In scanning the parameter space, we employ the Metropolis-Hastings
algorithm as described in \cite{Belanger:2009ti}. The data points
collected all satisfy
the requirement of REWSB,
with the neutralino in each case being the LSP. After collecting the data, we impose
the mass bounds on all the particles \cite{Nakamura:2010zzi} and use the
IsaTools package~\cite{Baer:2002fv}
to implement the various phenomenological constraints. We successively apply the following experimental constraints on the data that
we acquire from ISAJET:

\begin{table}[h]\centering
\begin{tabular}{rlc}
$m_h~{\rm (lightest~Higgs~mass)} $&$ \geq\, 114.4~{\rm GeV}$          &  \cite{Schael:2006cr} \\
$BR(B_s \rightarrow \mu^+ \mu^-) $&$ <\, 1.2 \times 10^{-8}$        &   \cite{:2007kv}      \\
$2.85 \times 10^{-4} \leq BR(b \rightarrow s \gamma) $&$ \leq\, 4.24 \times 10^{-4} \;
 (2\sigma)$ &   \cite{Barberio:2008fa}  \\
$0.15 \leq \frac{BR(B_u\rightarrow
\tau \nu_{\tau})_{\rm MSSM}}{BR(B_u\rightarrow \tau \nu_{\tau})_{\rm SM}}$&$ \leq\, 2.41 \;
(3\sigma)$ &   \cite{Barberio:2008fa}  \\
$\Omega_{\rm CDM}h^2 $&$ =\, 0.1123 \pm 0.0035 \;(5\sigma)$ &
\cite{Komatsu:2008hk} \\ $ 0 \leq \Delta(g-2)_{\mu}/2 $ & $ \leq 55.6 \times 10^{-10} $ & \cite{Bennett:2006fi}
\end{tabular}\label{table}
\end{table}

Employing the boundary condition from Eq.(\ref{gaugino10}) one  can define the MSSM gaugino masses at $ M_{\rm GUT} $ in terms of the mass parameter $M_{1/2}$ :
\begin{align}
M_1= M_{1/2} \nonumber \\
M_2= 3M_{1/2} \nonumber \\
M_3= - 2 M_{1/2}
 \label{gaugino11}
\end{align}
Note that $M_2$ and $M_3$ have opposite signs which, as we show, is important implementing  Yukawa coupling unification to a high accuracy.


\section{Threshold corrections and Yukawa unification\label{muneg}}

The SUSY threshold corrections to the top, bottom and tau Yukawa couplings
play a crucial role in $t$-$b$-$\tau$ Yukawa coupling unification.
In general, the bottom Yukawa coupling $y_b$ can receive large
threshold corrections, while the threshold corrections to $y_t$ are
typically smaller~\cite{Pierce:1996zz}. The scale at which Yukawa coupling unification
occurs is identified with $M_{\rm GUT}$, the scale of gauge coupling unification.
Consider first the case $y_t ( M_{\rm GUT})\approx y_{\tau}(M_{\rm
GUT})$. The SUSY correction to the
tau lepton mass  is given by $\delta m_{\tau}=v\cos\beta \delta y_{\tau}$.
For the large $\tan\beta$ values of interest here, there is sufficient
freedom in the choice of $\delta y_{\tau}$ to achieve $y_t\approx y_{\tau}$
at $M_{\rm GUT}$. This freedom stems from the fact that
$\cos\beta \simeq 1/\tan\beta$ for large $\tan\beta$, and so we may
choose an appropriate $\delta y_{\tau}$ and $\tan\beta$ to give us both
the correct $\tau$ lepton mass and $y_t \approx y_{\tau}$.
The SUSY contribution to $\delta y_b$ has to be carefully monitored
in order to achieve Yukawa coupling unification $y_t (M_{\rm
GUT})\approx y_{b}(M_{\rm GUT}) \approx y_{\tau}(M_{\rm GUT})$.

We choose the sign of  $\delta y_i\, (i=t,b,\tau)$ from the perspective of
evolving $y_i$ from $M_{\rm GUT}$ to $M_{\rm Z}$. With this choice,
$\delta y_b$ must receive a negative contribution
($-0.27 \lesssim \delta y_b/y_b \lesssim -0.15$) in order to realize
Yukawa coupling unification. This is a narrow interval considering
the full range of $-0.4\lesssim \delta y_b/y_b \lesssim 0.4$.
The dominant contribution to $\delta y_b$ arises from
the finite gluino and chargino loop corrections,
and in our sign convention, it is approximately given by~\cite{Pierce:1996zz}

\begin{align}
\delta y_b^{\rm finite}\approx\frac{g_3^2}{12\pi^2}\frac{\mu m_{\tilde g}
\tan\beta}{m_{\tilde b}^2}+
                         \frac{y_t^2}{32\pi^2}\frac{\mu A_t \tan\beta}{m_{\tilde t}^2},
\label{finiteCorrectionsEq}
\end{align}
where $g_3$ is the strong gauge coupling, $m_{\tilde g}$ is
the gluino mass, $m_{\tilde b}$ and $m_{\tilde t}$ are the
sbottom and stop masses, and $A_t$ is the
top trilinear (scalar) coupling.

The logarithmic corrections to $y_b$ are positive, which
leaves the finite corrections to provide for the correct overall negative
$\delta y_b$ in order to realize Yukawa unification. The gluino contribution
(in Eq.(\ref{finiteCorrectionsEq})) is positive for $\mu>0$ for the case of same sign
gaugino SBB mass term. Thus, the chargino
contribution (in Eq.(\ref{finiteCorrectionsEq})) must play an essential role
to provide the required negative contribution to $\delta y_b$. This can be
achieved with suitably large values of both $m_{16}$ and $A_t$,
which is the reason behind the requirement $m_{16}\gtrsim 6\,{\rm TeV}$ and $A_0/m_{16}\sim -2.6$ in the $SO(10)$ model
discussed in \cite{Baer:2008jn, Gogoladze:2009ug}.

One can improve the situation immensely by considering the case
of opposite sign gaugino soft terms which is allowed by the 4-2-2
model. Reference \cite{Gogoladze:2010fu} shows the parameter space, corresponding to $\mu< 0, M_2<0, M_3>0$, that gives Yukawa
coupling unification with a sub-TeV sparticle
spectrum, and which is consistent with all known experimental bounds
including $\Delta(g-2)_\mu$. Another possibility is to
consider $\mu>0,M_2>0,M_3<0$. This choice also improves the situation about $ t$-$b$-$\tau $ Yukawa unification, since it provides a negative gluino loop contribution to $ \delta y_{b} $.



\section{ $t$-$b$-$\tau$ Yukawa  unification with $m_{H_{u}}=m_{H_{d}}$\label{huhd}}

After scanning the parameter space listed in Eq.(\ref{parameterRange}),  and employing the GUT scale  boundary condition for the gauginos in Eq.(\ref{gaugino11}), we find that REWSB can occur, compatible with  $t$-$b$-$\tau$ Yukawa  unification, even if we set $m_{H_{u}}=m_{H_{d}}$. Note that this is very difficult
 to achieve if gaugino mass universality is imposed at $ M_{\rm GUT} $ \cite{Olechowski:1994gm}.
In order to explain our findings, consider the following one loop renormalization group equations for $ m^{2}_{H_{u}} $, $ m^{2}_{H_{d}} $ and the soft trilinear terms $ A_{t,b,\tau} $ \cite{Martin:1997ns}:
\begin{eqnarray}
16 \pi^2 \frac{d}{dt} m_{H_u}^2 \!&=&\!
3 X_t - 6 g_2^2 |M_2|^2 - \frac{6}{5} g_1^2 |M_1|^2 + \frac{3}{5} g^{2}_1,
\label{mhurge}
\\
16\pi^2 \frac{d}{dt} m_{H_d}^2 \!&=&\!
3 X_b + X_\tau - 6 g_2^2 |M_2|^2 - \frac{6}{5} g_1^2 |M_1|^2 - \frac{3}{5}
g^{2}_1 S.\label{mhdrge}
\end{eqnarray}

\begin{eqnarray}
16\pi^2 {d\over dt} A_t \!&=&\! A_t \Bigl [ 18 y_t^2  + y_b^2
- {16\over 3} g_3^2 - 3 g_2^2 - {13\over 15} g_1^2 \Bigr ]
+ 2 A_b y_b  y_t
\nonumber\\ &&
\!+ y_t \Bigl [ {32\over 3} g_3^2 M_3 + 6 g_2^2 M_2 + {26\over 15} g_1^2 M_1
\Bigr ],
\label{atrge}
\\
16\pi^2{d\over dt} A_b \!&=&\! A_b \Bigl [ 18 y_b^2 + y_t^2 +
y_\tau^2
- {16\over 3} g_3^2 - 3 g_2^2 - {7\over 15} g_1^2 \Bigr ]
+ 2 A_t y_t  y_b + 2 A_\tau y_\tau  y_b
\phantom{xxxx}
\nonumber \\&&
\!+ y_b \Bigl [ {32\over 3} g_3^2 M_3 + 6 g_2^2 M_2 + {14 \over 15} g_1^2 M_1
\Bigr ],\qquad{}
\\
16\pi^2{d\over dt} A_\tau \!&=&\! A_\tau \Bigl [ 12 y_\tau^2
+ 3 y_b^2 - 3 g_2^2 - {9\over 5} g_1^2 \Bigr ]
+ 6 A_b y_b  y_\tau
+ y_\tau \Bigl [ 6 g_2^2 M_2 + {18\over 5} g_1^2 M_1 \Bigr ].
\label{brge}
\end{eqnarray}

Here
\begin{eqnarray}
X_t  &=  & 2 |y_t|^2 (m_{H_u}^2 + m_{\tilde Q_3}^2 + m_{\tilde u_3^c}^2) +2 |A_t|^2,
\label{xt}
\\
X_b  &= & 2 |y_b|^2 (m_{H_d}^2 + m_{\tilde Q_3}^2 + m_{\tilde d_3^c}^2) +2 |A_b|^2,
\label{xb}
\\
X_\tau  &= &  2 |y_\tau|^2 (m_{H_d}^2 + m_{\tilde L_3}^2 + m_{\tilde e_3^c}^2)
+ 2 |A_\tau|^2,
\label{xtau}
\\
S &\equiv & {\rm Tr}[Y_j m^2_{\phi_j}] =
m_{H_u}^2 - m_{H_d}^2 + {\rm Tr}[
{ m^2_{\tilde Q}} - { m^2_{\tilde L}} - 2 { m^2_{{\tilde u}^c}}
+ { m^2_{{\tilde d}^c}} + { m^2_{{\tilde e}^c}}] ,
\label{eq:defS}
\end{eqnarray}
and ${\tilde Q_3}$, ${\tilde u_3^c}$, $\tilde d_3^c$, ${\tilde L_3}$, ${\tilde e_3^c}$ denote the third generation squarks and sleptons.
Also, $g_i$ and $M_i$  ($i=1,2,3$) denote the gauge couplings and gaugino masses
for $U(1)_Y$, $SU(2)_L$ and $SU(3)_C$, and $y_j$, $A_j$ ($j=t, b, \tau$) are the third family Yukawa couplings and trilinear scalar SSB couplings, respectively.

In Eqs. (\ref{mhurge}) and (\ref{mhdrge}), terms with negative coefficients drive $ m^{2}_{H_{u}} $ and $ m^{2}_{H_{d}} $ to larger values in the evolution from $ M_{\rm GUT} $ to $ M_{Z} $, while the terms with positive signs clearly have the opposite effect.  On the other hand, at scale $M_Z$ ,the electroweak symmetry breaking minimization condition at tree level requires that
\begin{eqnarray}
\mu^2=\frac{m_{H_d}^2 - m_{H_u}^2\tan^{2}\beta}{\tan^{2}\beta -1} - \frac{M^{2}_{Z}}{2}
\end{eqnarray}

For moderate to large values for $\tan\beta$ and assuming that $|m_{H_u}^2 |\gg M_Z^2$, we can get $ \mu^{2}\approx -m^{2}_{H_{u}} $. In order to have REWSB, $ m^{2}_{H_{u}} $ must be driven to negative values. In the literature, two alternative ways  are used to achieve  $m_{H_u}^2<0$.
In one case  with $y_t>y_b,~y_{\tau}$, $m^{2}_{H_{u}}$ is driven to negative values at $ M_{Z} $ by $y_t$, as seen from Eqs.(\ref{mhurge}) and (\ref{xt}). In this case it is possible to find solutions for which $m_{H_{u}}=m_{H_{d}}$ at $ M_{\rm GUT} $. In a model with universal gaugino masses at $ M_{\rm GUT} $, we need  a very large $m_{16}>6$ TeV and and even larger trilinear SSB coupling, $A_0\sim - 2.6 m_{16}$, in order that $y_t(M_{\rm GUT})=y_b(M_{\rm GUT})= y_{\tau}(M_{\rm GUT})$. As we know, one has $y_t>y_b>y_{\tau}$ all the way from $ M_{\rm GUT} $ to $M_Z$, and the SSB terms in Eqs.(\ref{xt}) and (\ref{xb}) renormalize more or less equally. This means that $X_t>X_b$, but the additional contribution from $X_{\tau}$ makes the contributions from the positive terms in  Eq.(\ref{mhdrge}) larger than the corresponding positive contributions in Eq. (\ref{mhurge}).  As a result, $m^{2}_{H_{d}}$ is driven to negative values earlier than $m^{2}_{H_{u}}$
 at $ M_{Z} $, which is phenomenologically not acceptable. To avoid this problem,  one can impose $m_{H_{u}}<m_{H_{d}}$ at $ M_{\rm GUT} $.  As we mentioned in Section \ref{muneg}, with opposite signs for $M_2$ and $M_3$ at $ M_{\rm GUT} $, $t$-$b$-$\tau$ Yukawa unification becomes easier by appropriately choosing the sign for $ \mu $ in  Eq. (\ref{finiteCorrectionsEq}) \cite{Gogoladze:2009ug, Gogoladze:2010fu}, and one can have $t$-$b$-$\tau$ Yukawa unification even for $m_{16}\sim 1-2$ TeV. Notice that the relative sign of $M_2$ and $M_3$ affects the RG running of the trilinear SSB couplings in Eqs.(\ref{atrge})-(\ref{brge}). In most cases, we find $ A_{t}\sim A_{b}$, while $A_b$ and $A_{\tau}$ have opposite sign at $M_Z$.
{We see from Figure 1a that $A_{\tau}$ is driven towards zero and then continues its evolution.
 This tendency helps in changing the direction of evolution for $m^2_{H_u}$.}

As we will show in the next section, for precise $t$-$b$-$\tau$ Yukawa coupling unification, $m_{H_{u}}=m_{H_{d}}= m_{10}<1$ TeV at $ M_{\rm GUT} $. Hence, we can conclude that the SSB terms in Eqs.(\ref{xt}) and (\ref{xb}) are comparable. Since the values of the SSB terms for opposite signs of gaugino masses are around  1 TeV, the contribution from  $X_{\tau}$ in Eq. (\ref{mhdrge}) is not very significant. The coupling $y_t$ is slightly larger than $y_b$ during RG running, but in our case it is enough to drive $ m_{H_u}^2$ negative at the weak scale.

\begin{figure}[t!]
\centering
\includegraphics[width=15.5cm]{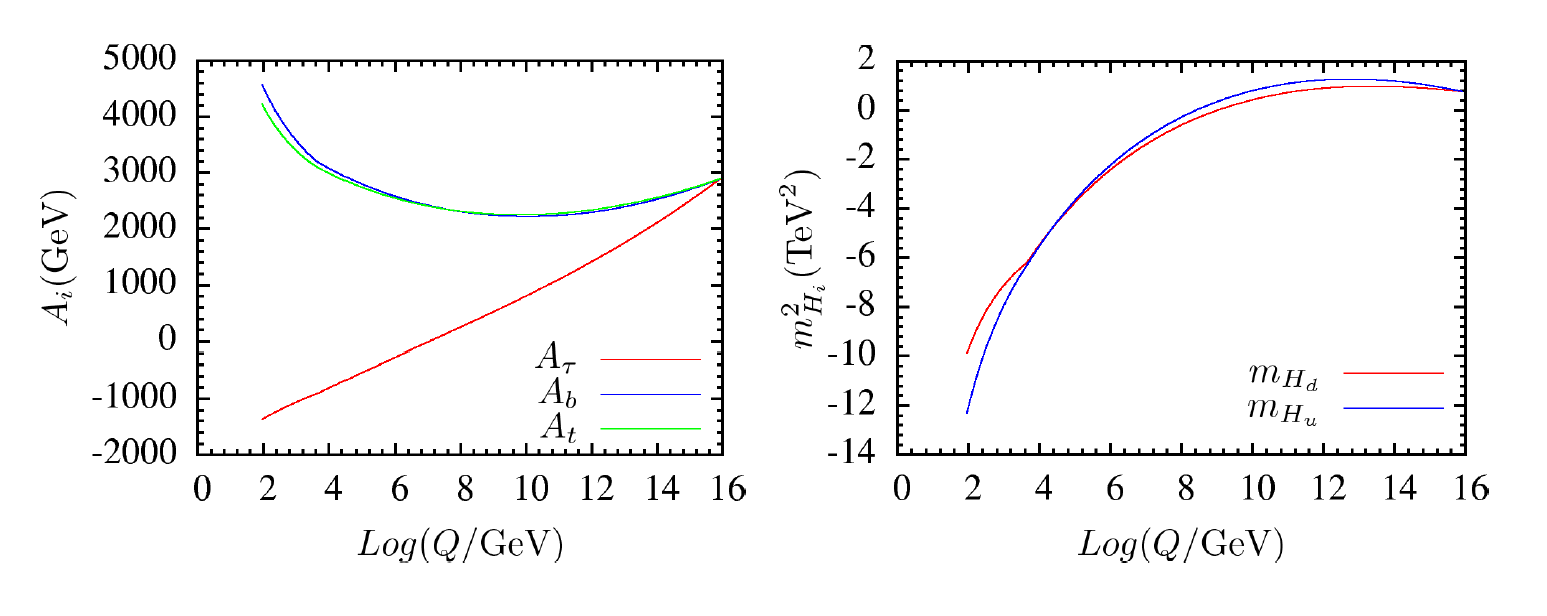}
\caption{ Evaluation of trilinear  SSB couplings (left panel) and the MSSM Higgs (right panel)  SSB   mass terms corresponding to the benchmark point 1 in Table 1.}
\label{fund-1}
\end{figure}

The interplay among these various contributions is neatly summarized in Figure \ref{fund-1}. In the left panel we plot the RG evolution of the three soft trilinear couplings, while the right panel shows the corresponding evolution of $ m^{2}_{H_{u}} $ and $ m^{2}_{H_{d}} $.
{In order to have phenomenologically acceptable EWSB we need to have $m^2_{H_u}$ more negative than $m^2_{H_d}$ at EWSB scale. The relation at high scale between $ m^{2}_{H_{u}} $ and $ m^{2}_{H_{d}} $ is not very  important.  Figure 1b shows that at the EWSB scale $m^2_{H_u}$ is more negative than $m^2_{H_d}$.
In the text we use $M_Z$ as the scale of EWSB.}

\section{Higgs Boson Mass and Sparticle Spectroscopy \label{results}}

 \begin{figure}[t!]
\centering
\includegraphics[width=13.cm]{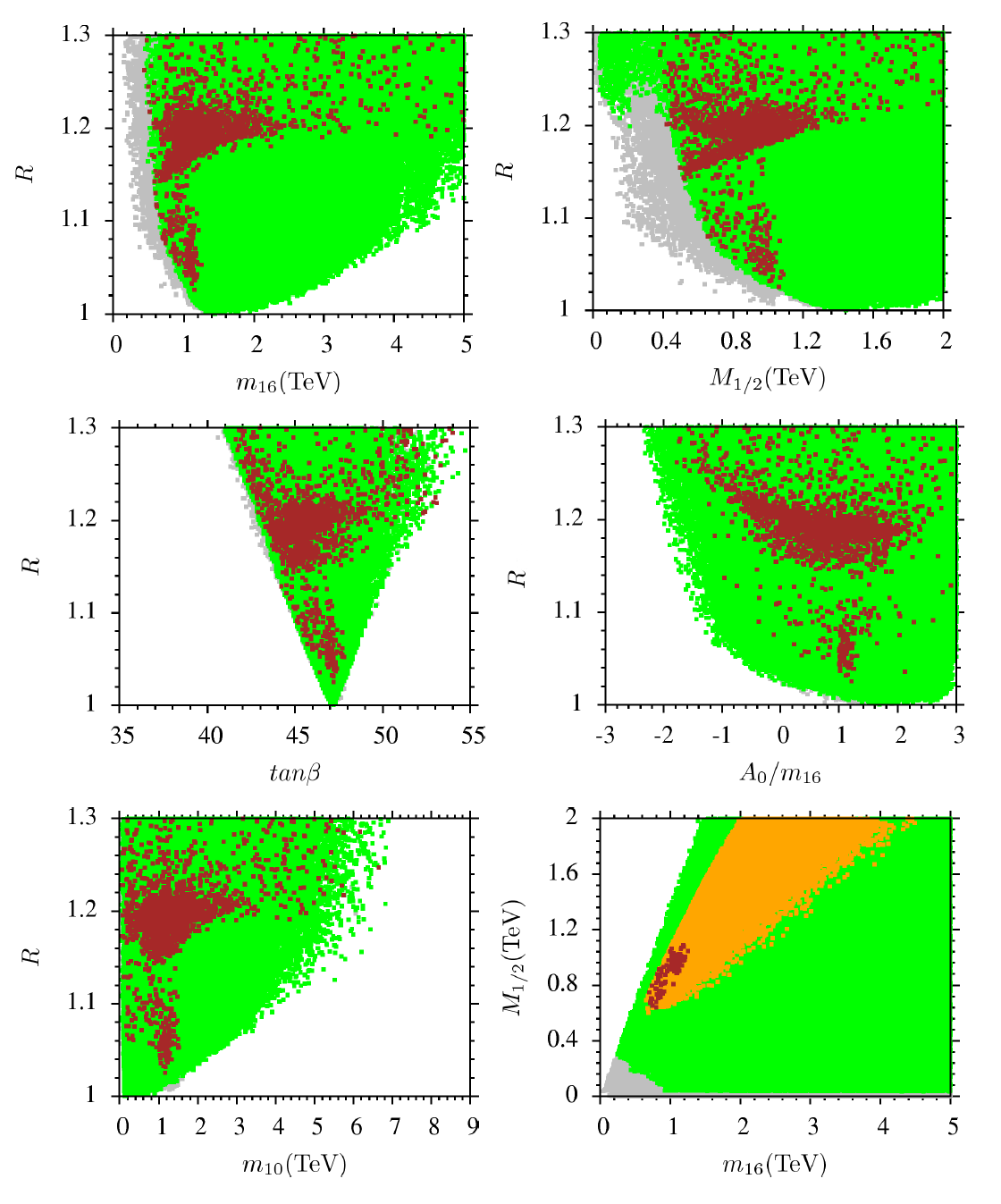}
\caption{Plots in $R-m_{16}$,  $R-M_{1/2}$, $R-\tan\beta$, $ R-A_{0}/m_{16} $, $ R-m_{10} $ and $ M_{1/2}-m_{16} $ planes. Panels correspond to the following choice of parameters: $\mu > 0$, $M_2 > 0$ and $M_3:M_2:M_1=-2:3:1$. Gray points are consistent with REWSB and neutralino LSP. Green points satisfy particle mass bounds and constraints from $BR(B_s\rightarrow \mu^+ \mu^-)$, $BR(b\rightarrow s \gamma)$
and $BR(B_u\rightarrow \tau \nu_\tau)$. In addition, we require that green points do no worse than the SM in terms of $(g-2)_\mu$. Brown  points belong to a subset of green points and satisfy the WMAP bounds on neutralino dark matter abundance.}
\label{fund-2}
\end{figure}

\begin{figure}[t!]
\centering
\includegraphics[width=8.cm]{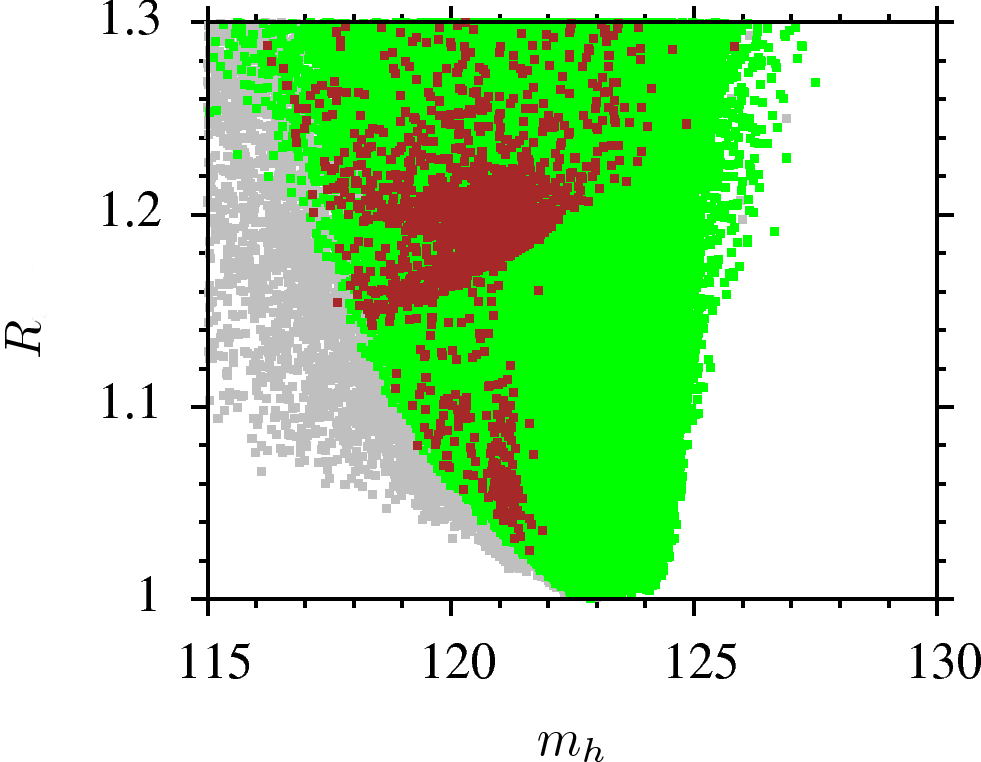}
\caption{Plot in $R-m_h$ plane. The color coding is the same as that used in Fig. \ref{fund-2}.}
\label{fund-3}
\end{figure}

\begin{figure}[t!]
\centering
\includegraphics[width=15.cm]{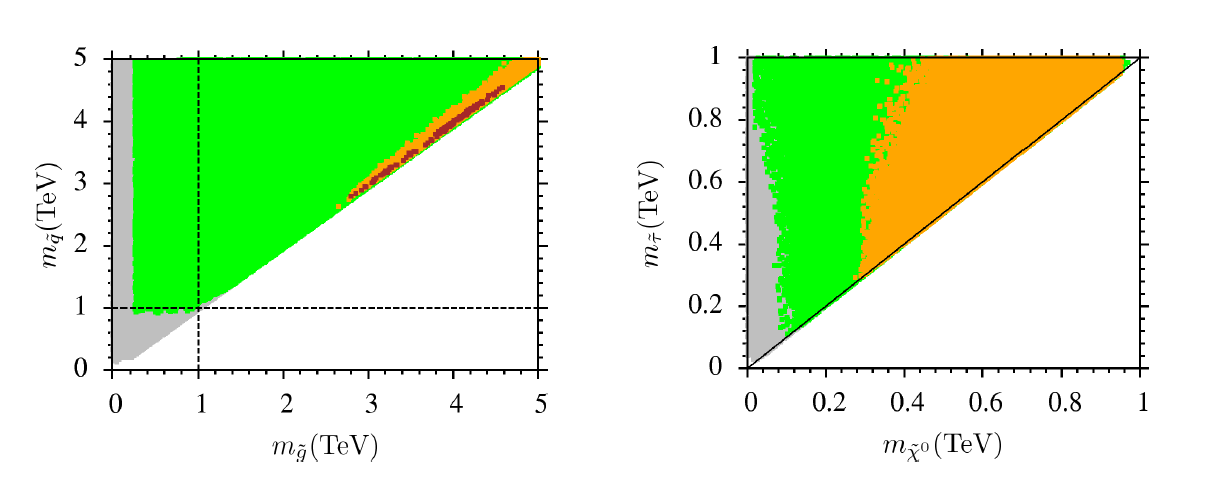}
\caption{Plots in $m_{\tilde{q}}-m_{\tilde{g}}$, $m_{\tilde{\tau}}-m_{\tilde{\chi}^0}$ planes. The color coding is the same as that used in Fig. \ref{fund-2}. In addition, yellow points represent Yukawa unification within to $ 10\% $. Brown points in $ m_{\tilde{q}}-m_{\tilde{g}} $ plane that are compatible with WMAP bound on relic abundance are a subset of yellow points in $ m_{\tilde{q}}-m_{\tilde{g}} $ plane.}
\label{fund-4}
\end{figure}

In order to quantify Yukawa coupling unification, we define the quantity $R$ as,
\begin{align}
R=\frac{ {\rm max}(y_t,y_b,y_{\tau})} { {\rm min} (y_t,y_b,y_{\tau})}
\end{align}
{In Figure ~\ref{fund-2}  we show the
results in the  $R-m_{16}$,  $R-M_{1/2}$, $R-\tan\beta$, $ R-A_{0}/m_{16} $, $ R-m_{10} $ and $ M_{1/2}-m_{16} $ planes. The panels correspond to the following choice of parameters: $\mu > 0$, $M_2>0$ and $M_3:M_2:M_1=-2:3:1$.} The gray points are consistent with REWSB and neutralino LSP.
The green points satisfy the particle mass bounds and constraints from $BR(B_s\rightarrow \mu^+ \mu^-)$, $BR(b\rightarrow s \gamma)$ and $BR(B_u\rightarrow \tau \nu_\tau)$. In addition, the green points do no worse than the SM in terms of $(g-2)_\mu$. The brown  points belong to a subset of the green points and satisfy
the WMAP bounds on neutralino dark matter abundance.

In the $R$ - $m_{16}$ plane of Figure~\ref{fund-2} {we see that we can realize
perfect Yukawa unification consistent with all constraints mentioned in
Section~\ref{constraintsSection}.} This is possible because  we can implement Yukawa
unification for relatively small values of $m_{16}$ ($\sim 1\, {\rm TeV}$). This is more than an
order of magnitude reduction in the $m_{16}$ values required for
Yukawa unification, compared with the case $\mu>0$ and universal gaugino masses. In  the $R- M_{1/2}$ plane of Figure ~\ref{fund-2},  we see that employing the boundary conditions for gauginos presented in Eq.  (\ref{gaugino11}),
perfect $t$-$b$-$\tau$ Yukawa  unification prefers heavier ($> 1.2$ TeV) values for $M_{1/2}$. We can also predict that $\tan\beta \approx 47$. In this case for Yukawa unification up to 5$\%$ level, $45\leq\tan\beta\leq48$.

{In the $R-A_0/m_{16}$ plane of Figure~\ref{fund-2} we see that in the present $ SO(10) $ case with 10$\%$ or better  $t$-$b$-$\tau$  Yukawa  unification, we obtain a relaxation of $A_0$ values  similar to the $ SU(4)_{c}\times SU(2)_{L}\times SU(2)_{R} $ model in  \cite{Gogoladze:2010fu}, with  $-2.5<A_0/m_{16}<1$ for $\mu<0$, and  $-1.5<A_0/m_{16}<3$ for $\mu>0$.  The $R-m_{10}$ plane shows that precise unification prefers 0.2 TeV $<m_{10}< 1$ TeV. We see also that relaxing the Yukawa unification condition relaxes the upper bound for $m_{10}$.}
 In order to better visualize the magnitude of the sparticle masses consistent with $t$-$b$-$\tau$ Yukawa unification, we present our results in the $M_{1/2}-m_{16}$ plane, where the yellow points correspond to Yukawa unification better than 10$\%$.

{In Figure ~\ref{fund-3}  we show the
results in the $R-m_h$ plane. It is most interesting to note that demanding precise $t$-$b$-$\tau$ Yukawa unification for $\mu>0$, $M_2>0$ and $M_3<0$ allows us to predict the light CP even Higgs mass in a very narrow interval, 122 GeV $<m_h<$ 124 GeV. For Yukawa unification of up to $5\%$, it becomes   120 GeV $<m_h<$ 125 GeV. We expect a theoretical uncertainty of about 3 GeV in the calculation of the light Higgs mass.}

{In Figure \ref{fund-4} the $m_{\tilde{q}}-m_{\tilde{g}}$ panel shows that $t$-$b$-$\tau$ Yukawa unification in our scenario predicts masses for the gluino and the first two family squarks which lie somewhat beyond the current  ATLAS \cite{Aad:2011ib} and CMS \cite{Chatrchyan:2011zy} bounds.}

{In the present framework, the WMAP constraint on the relic dark matter abundance is only satisfied by the neutralino-stau co-annihilation scenario.
 We know that in neutralino-stau coannihilation, the neutralino and stau masses are degenerate to a good approximation.
Instead of introducing a new color in the figure, we decided just to draw the unit line, and then it is easy to understand that points
close to this line will give good relic abundance through neutralino-stau coannihilation.
 From the plot in $ m_{\tilde{\tau}}-m_{\tilde{\chi}^{0}} $ plane of Figure \ref{fund-4}, we see that the LSP neutralino mass is greater than or of order 300 GeV. Other co-annihilation scenarios may emerge after doing a more thorough analysis, but this remains to be seen.}

Finally, in Table 1 we present some benchmark points with $\mu>0$, $M_2>0$ and $M_3<0$.
All of these benchmark points satisfy the various constraints, except possibly for a small discrepancy with the WMAP bound on relic dark density. Points 1 and 2 depict solutions corresponding to minimum R and best $ \Omega_{CDM} h^{2} $ values. Point 2 describes a solution with $m_{16}=m_{10}$.  Point 3 displays a solution with a Higgs boson mass close to 124 GeV. The theoretical uncertainty in this estimate is 2-3 GeV.

\begin{table}[t!]
\hspace{-1.0cm}
\centering
\begin{tabular}{|c|ccc|}
\hline
\hline
                 & Point 1 & Point 2 & Point 3\\
\hline
$m_{16}$         & 1277 & 1296 & 1920\\
$M_{1} $         & 1302 & 1174 & 1768\\
$M_{2} $         & 3906 & 3522 & 5304\\
$M_{3} $         & -2604 & -2248 & -3536\\
$m_{10}$         & 613.7 & 1296 & 240.2\\
$\tan\beta$      & 47.2 & 47.3 & 46.9\\
$A_0/m_{16}$        & 1.41 & 1.61 & 2.73\\
$m_t$            & 173.1  & 173.1 & 173.1\\
\hline
$\mu$            & 2216 & 1919.7 & 3478.7\\
$ B_{\mu} $      & 20.3 & 19.5 & 26.9 \\

\hline
$m_h$            & 122.3 & 122.1 & 124\\
$m_H$            & 408.7 & 392.5 & 559.5\\
$m_A$            & 406 & 390 & 555.8\\
$m_{H^{\pm}}$    & 420.5 & 404.7 & 568.3\\

\hline
$m_{\tilde{\chi}^0_{1,2}}$
                 & 609.3, 2227.3 & 546.7, 1930.4 & 833.4, 3487.6\\
$m_{\tilde{\chi}^0_{3,4}}$
                 & 2230.4, 3314.7 & 1933.7, 2986.2 & 3490.7, 4503.2\\

$m_{\tilde{\chi}^{\pm}_{1,2}}$
                 & 2260.4, 3281.1 & 1959.8, 2956 & 3533.3, 4458.5\\
$m_{\tilde{g}}$  & 5413.9 & 4919.5 & 7217.9\\

\hline $m_{ \tilde{u}_{L,R}}$
                 & 5251.4, 4766.1 & 4895.8, 4375.4 & 7176.2, 6386\\
$m_{\tilde{t}_{1,2}}$
                 & 3839.2, 4665.7 & 3453.1, 4211.6 & 4967.6, 6120.1\\
\hline $m_{ \tilde{d}_{L,R}}$
                 & 5352, 4759.7 & 4896.5, 4369.9 & 7176.7, 6378.7\\
$m_{\tilde{b}_{1,2}}$
                 & 3969.3, 4640.1 & 3589.8, 4188 & 5168.1, 6088.1\\
\hline
$m_{\tilde{\nu}_{1}}$
                 & 2784.8 & 2585 & 3844.3\\
$m_{\tilde{\nu}_{3}}$
                 &  2639.9 & 2421.7 & 3605.8\\
\hline
$m_{ \tilde{e}_{L,R}}$
                & 2789.5, 1355.5 & 2589.3, 1359.5 & 3849.1, 2016.8\\
$m_{\tilde{\tau}_{1,2}}$
                & 609.4, 2650.7 & 548.9, 2431.6 & 842.4, 3621.2\\
\hline

$\Delta(g-2)_{\mu}$  & $0.68\times 10^{-10} $ & $ 0.84\times 10^{-10} $ & $0.35\times 10^{-10} $\\

$\sigma_{SI}({\rm pb})$
                & $0.37\times 10^{-9}$ & $ 0.59\times 10^{-9} $ & $0.46\times 10^{-10}$\\

$\sigma_{SD}({\rm pb})$
                & $0.47\times 10^{-8}$ & $ 0.89\times 10^{-8} $ & $0.63\times 10^{-9}$\\

$\Omega_{CDM}h^{2}$
                &  0.176 & 0.189 & 0.55\\
\hline

$R$     &1.01 & 1.02 & 1.01\\

\hline
\hline
\end{tabular}
\vspace{0.5cm}
\caption{Sparticle and Higgs masses (in GeV) for $\mu>0$.   Points 1 and 2 depict solutions corresponding to minimum R and best $ \Omega_{CDM} h^{2} $ values. Point 2 corresponds to the case when $m_{16}=m_{10}$. Point 3 displays a solution with the heaviest  mass for CP even light Higgs boson compatible with Yukawa unification better than 1$\%$.}
\end{table}


\section{Conclusion \label{conclusions}}

We have reconsidered $t$-$b$-$\tau$ Yukawa unification within a slightly revised framework in this paper. The main difference from most previous investigations stems from the assumptions we make related to the soft supersymmetry breaking parameters. First, we assume that the MSSM gauginos have non-universal masses, which are related to one another at  $M_{\rm GUT}$ by some appropriate SO(10) group theory factors. Second, we set the masses of the two MSSM Higgs doublets to be equal at $M_{\rm GUT}$. Overall, this means that we effectively have one parameter less than in the standard approach to SO(10) Yukawa unification.

The ramifications of these slightly different assumptions for TeV scale physics turn out to be quite startling, with the low energy predictions being very different from previous studies. We find, for instance, that $t$-$b$-$\tau$ Yukawa unification solutions at 5$\%$  level or better exist in our model with $\mu > 0$ and $ M_{2} > 0 $.

These solutions, obtained using the ISAJET  software, are compatible with all experimental observations, as well as the WMAP dark matter constraint (through stau co-annihilation). The masses of the gluino and first two family squarks are found to lie in the 2.7 - 5 TeV range, while the lightest stop (top squark) weighs at least 2 TeV or so. Finally, with 5$\%$ or better Yukawa unification, the lightest Higgs boson is predicted to have a mass of around 120 - 125 GeV mass range, (with an uncertainty of $\pm 3$ GeV). The MSSM parameter $\tan\beta$ is around 45 - 47.

\section*{Acknowledgments}
We thank M.~Adeel Ajaib, Azar Mustafayev and Shabbar Raza  for valuable discussions.
This work is supported in part by the DOE Grant No. DE-FG02-91ER40626
(I.G., Q.S. and C.S.U.). This work used the Extreme Science
and Engineering Discovery Environment (XSEDE), which is supported by the National Science
Foundation grant number OCI-1053575.


\end{document}